\documentclass[aps,prd,floatfix,twocolumn,nofootinbib,superscriptaddress,longbibliography]{revtex4-2}

\usepackage{graphicx}

\usepackage{todonotes}
\usepackage{float}
\usepackage{orcidlink}
\usepackage{dcolumn}
\usepackage{bm}
\usepackage{hyperref}
\hypersetup{
   breaklinks=true,   
}
\begin{document}
\preprint{APS/123-QED}
\title{Enhanced antineutrino emission from $\beta$~decay in core-collapse supernovae with self-consistent weak decay rates}%

\author{T. Dasher\,\orcidlink{0009-0001-4814-2185}}
\email{dashertr@msu.edu}
\affiliation{Facility for Rare Isotope Beams, Michigan State University, East Lansing, Michigan 48824, USA}
\affiliation{Department of Physics and Astronomy, Michigan State University, East Lansing, Michigan 48824, USA}

\author{A.~Ravli\'c\,\orcidlink{0000-0001-9639-5382}}
\email{ravlic@frib.msu.edu}
\affiliation{Facility for Rare Isotope Beams, Michigan State University, East Lansing, Michigan 48824, USA}
\affiliation{Department of Physics, Faculty of Science, University of Zagreb, Bijeni\v cka c. 32, 10000 Zagreb, Croatia}

\author{S.~Lalit\,\orcidlink{0000-0001-7758-492X}}
\email{lalit@frib.msu.edu}
\affiliation{Facility for Rare Isotope Beams, Michigan State University, East Lansing, Michigan 48824, USA}

\author{E.~O'Connor\,\orcidlink{0000-0002-8228-796X}}
\email{evan.oconnor@astro.su.se}
\affiliation{The Oskar Klein Centre, Department of Astronomy, Stockholm University, AlbaNova, SE-106 91 Stockholm, Sweden}

\author{K.~Godbey\,\orcidlink{0000-0003-0622-3646}}
\email{godbey@frib.msu.edu}
\affiliation{Facility for Rare Isotope Beams, Michigan State University, East Lansing, Michigan 48824, USA}

\date{\today}%

\begin{abstract}
Nuclear weak-interaction rates are known to exert a prominent effect in the late-stages of stellar collapse. Despite their importance, most studies to date on core-collapse supernovae (CCSNe) have focused primarily on the effects of electron captures, neglecting $\beta$~decay contributions. 
In this work, we present the first CCSNe simulation incorporating global $\beta$~decay rates from a microscopic theory. These are enabled by a large-scale evaluation of both electron capture and $\beta$~decay rates, obtained self-consistently utilizing the relativistic energy density functional theory and finite-temperature quasiparticle random-phase approximation. Including $\beta$ decay leads to a dramatic enhancement of the pre-bounce antineutrino signal as the antineutrino emissivity increases by more than two orders of magnitude and the luminosity by a factor of 50 relative to thermal emission alone, while the average antineutrino energy increases by over 1 MeV. It is expected that these new rates could help us constrain the model uncertainties related to weak-interaction processes, improving the prediction of antineutrino signal during the final stages of stellar death. 
\end{abstract}

\maketitle

\section{Introduction}
 The detection of a total of 25 events in the neutrino burst of supernova 1987A~\cite{Hirata1987,Bionta1987,Alexeyev1988} has led to several improvements in the detection techniques of the Earth-based neutrino observatories~\cite{Beacom2004,Odrzywolek2004, Odrzywolek2004b, Asakura2016,ProtoCollaboration2018} as well as global network for supernova early warning systems~\cite{Antonioli2004,AlKharusi2021, Kara2024} integrating presupernova antineutrino ($\bar{\nu}_e$) detection. 
Weak interactions play a central role in the final stages of stellar evolution, where stellar interiors are characterized by high temperatures and densities~\cite{Wilson1985,Langanke2021,RevModPhys.75.819,SUZUKI2022103974,Janke2007}. One of the most prominent processes during the early collapse phase of core-collapse supernovae (CCSNe) is electron capture (EC), with rates increasing rapidly at higher temperatures and densities. In contrast, due to blocking of the lepton phase-space, $\beta$~decay rates decrease considerably at higher densities. Nevertheless, during a brief but important interval in the presupernova phase, EC and $\beta$~decay can compete~\cite{MartinezPinedo2000}. At presupernova-like conditions, both EC and $\beta$~decays are dominated by allowed Fermi and Gamow-Teller (GT) transitions with nuclear composition consisting mostly of $pf$-shell nuclei~\cite{Bethe1990}. However, as the collapse proceeds, the contribution of forbidden transitions and heavier nuclei becomes non-negligible. EC and $\beta$~decay rates were initially estimated in Refs.~\cite{Fuller1980, Fuller1982a, Fuller1982b,Fuller1985} using the independent particle model with a single GT resonance, and supplemented, where available, by experimental data for low-lying transitions. Later, these rates were calculated using the nuclear shell-model (SM) up to $A=40$~\cite{Oda1994} and for $A=45-65$~\cite{Langanke1999,LANGANKE2000481,Zha2019}. Impact of $\beta$~decay rates on presupernova (anti)neutrinos, based on SM calculations for $pf$-shell nuclei, was investigated in Refs.~\cite{Patton2017a,Patton2017b,Heger2001a,Heger2001b}, revealing a substantial effect on the subsequent stellar evolution. Specifically, Ref.~\cite{Kato_2017} studied neutrino emission from the presupernova evolution through the pre-bounce collapse phase using tabulated EC and $\beta$ decay rates from several nuclear models where available. Although their reaction network included a broad set of nuclei, the microscopic rate tables covered only limited regions of the nuclear chart, while for nuclei outside these tables, approximate prescriptions were employed. While nuclei in the vicinity of ${}^{56}$Fe dominate the presupernova composition, heavier nuclei $(A > 60)$ increasingly contribute as the collapse proceeds~\cite{RevModPhys.75.819}. However, SM calculations quickly become numerically infeasible for these nuclei due to prohibitive scaling with the system size. For EC, this led to development of hybrid models~\cite{JUODAGALVIS2010454}, to cover the larger areas of nuclide chart. Still, $\beta$-decay calculations remain limited to SM. The approaches based on the energy density functional (EDF) theory, in principle, allow for extending calculations of weak-interaction processes throughout the nuclide chart, and also include more complicated forbidden transitions. Excited states can be efficiently probed employing the quasiparticle random-phase approximation (QRPA), with the residual interaction derived self-consistently from the underlying EDF~\cite{Paar2007}. This approach also allows for inclusion of finite-temperature effects within the finite-temperature QRPA (FT-QRPA)~\cite{Paar2009,PhysRevC.102.065804}. In particular, recent calculations involving consistent treatment of GT transitions in hot nuclei, within the thermal QRPA, predicted enhancement of the energy luminosity and average energies in ${}^{56}$Fe of the emitted (anti)neutrinos~\cite{Dzhioev2025,Dzhioev2023,Dzhioev2023b,Dzhioev2024}. In this work, following the recent work on EC in Ref.~\cite{Ravlic2025}, we perform large-scale calculations of $\beta$~decay rates obtained within the EDF+FT-QRPA framework and study its impact on the CCSNe evolution.

In Sec.~\ref{sec:theory} we present the theoretical framework employed to calculate $\beta$ decay rates, including the competition between EC and $\beta$ decay rates, based on the FT-QRPA calculations, in Sec.~\ref{sec:competition}, and comparison between FT-QRPA and SM rates in Sec~\ref{sec:comparison}. Results on CCSNe simulations for $15 M_\odot$ and $20 M_\odot$ progenitors are shown in Sec.~\ref{sec:results}, together with detectability analysis of the increased antineutrino signal.

\section{Theoretical framework}\label{sec:theory}
The theoretical framework employed for calculating the $\beta$~decay rates is based on the relativistic EDF theory with the momentum-dependent D3C$^*$ interaction \cite{PhysRevC.105.055801,PhysRevC.71.064301,PhysRevC.93.025805}, computational framework being detailed in Refs. \cite{Ravlic2025a,PhysRevC.105.055801}. The initial nuclear states are determined by the finite-temperature Hartree-Bardeen-Cooper-Schrieffer (FT-HBCS) theory assuming spherical symmetry \cite{PhysRevC.102.065804,PhysRevC.96.024303}, while the final states and transition strengths are determined within the FT-QRPA \cite{PhysRevC.102.065804,PhysRevC.101.044305,PhysRevC.104.054318}. For $\beta$~decays we carefully consider the contribution of de-excitations from highly-excited states in the parent nucleus~\cite{Ravlic2025a}. Apart from the allowed GT transitions, we also consider the first-forbidden (FF), $J^\pi = 0^-,\, 1^-$ and $2^-$, transitions that contribute to the $\beta$~decay rate for neutron-rich nuclei. Benchmarks of this model with experimental data on $\beta$ decay half-lives can be found in Ref.~\cite{Ravlic2021}. We note that this work uses an improved finite-range pairing force as discussed in Ref.~\cite{Ravlic2025}, which is more suitable for global calculations while preserving the level of agreement with experiment achieved in the previous implementation. Nuclei are assumed to be fully ionized with electrons (positrons) in the plasma described by the Fermi-Dirac distribution, meaning that $\beta$~decay rates are a function of the temperature $T$ and a product of stellar density with the electron-to-baryon ratio $\rho Y_e$. The $\beta$~decay rates can be expressed through the electron antineutrino spectral functions $n(E_{\bar{\nu}_e})$, summed over all the contributing multipoles $J^\pi$ \cite{PhysRevC.64.055801}
\begin{equation}\label{eq:beta_rate}
    \lambda_{\beta}^{(i)} = \sum \limits_{J^\pi}\int \limits_0^{Q_\beta} n_i(E_{\bar{\nu}_e}, J^\pi) d E_{\bar{\nu}_e},
\end{equation}
along with the antineutrino energy loss rate
\begin{equation}
    \lambda_{\bar{\nu}_e}^{(i)} = \sum \limits_{J^\pi}\int \limits_0^{Q_\beta} E_{\bar{\nu}_e}n_i(E_{\bar{\nu}_e}, J^\pi) d E_{\bar{\nu}_e},
\end{equation}
where $E_{\bar{\nu}_e}$ is the electron antineutrino energy, and $Q_\beta$ is the $\beta$~decay energy window. 

\begin{figure*}[t!]
    \centering
    \includegraphics[width=0.9\linewidth]{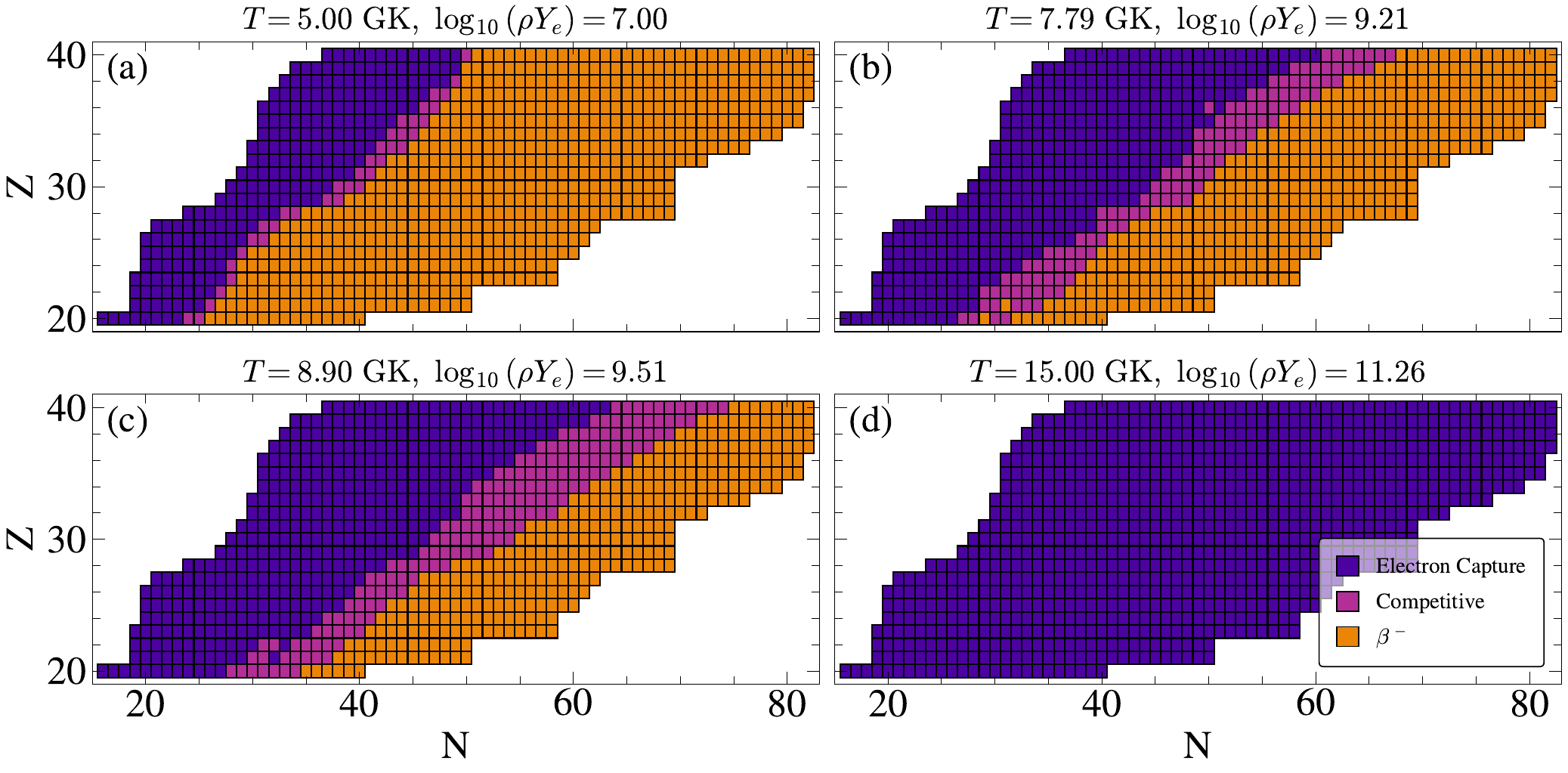}
    \caption{Competition between EC and $\beta$~decay rates for representative conditions during the CCSN collapse. Both sets of rates are calculated within the FT-QRPA framework, with EC rates in Ref.~\cite{Ravlic2025}.}\label{fig:S1}
\end{figure*}

\begin{figure*}[htbp!]
    \centering
     \includegraphics[width=0.6\linewidth]{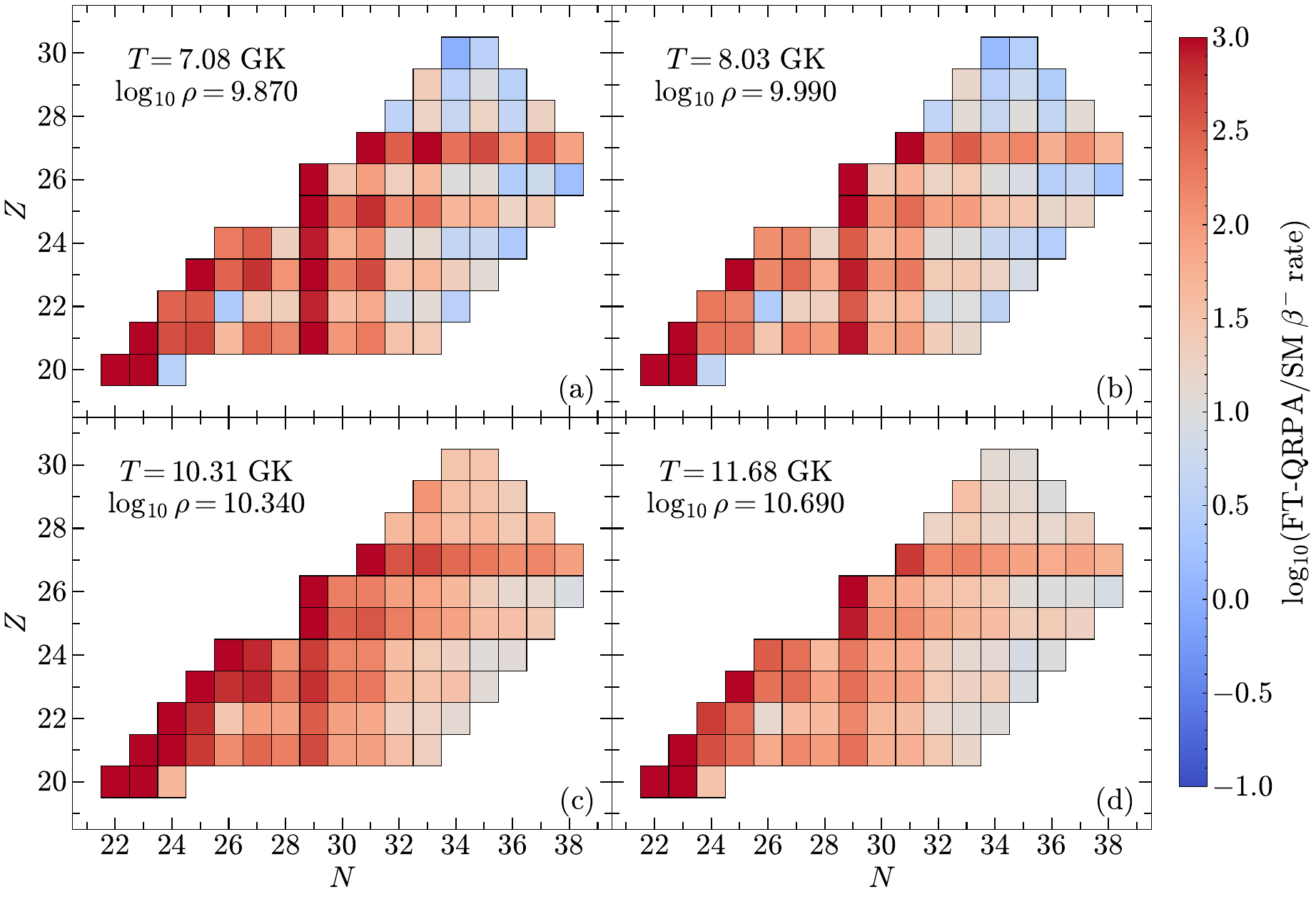}
     \caption{Comparison between our FT-QRPA and shell-model $\beta$ decay rates~\cite{Zha2019} at representative points along the CCSNe trajectory.}\label{fig:S2}
\end{figure*}

Model calculations include a large set of $\beta$~decay rates consisting of a total of 902 individual nuclei from the proton to the neutron drip line, and from calcium $(Z = 20)$ to zirconium $(Z = 40)$. The considered temperatures span a range from 1 to 30 GK, with a respective range in densities from $\rho Y_e = 10^5$ to $10^{12}$ g/cm${}^3$, enough to consider the CCSNe trajectories where $\beta$~decays are of relevance.
Of course, nuclei remain present at higher densities. However, beyond neutrino trapping between $\rho Y_e = 10^{11}$ to $10^{12}$ g/cm${}^3$ the net reaction rate is determined more based on the Pauli-blocking factor rather than the intrinsic nuclear structure effects. Additionally, $\beta$~decays are blocked at higher densities due to the Pauli-exclusion principle, as demonstrated in Fig.~\ref{fig:S1}. Thus, the inclusion of $\beta$~decays at higher densities is unnecessary.

The $\beta$~decay table includes a set of $(\lambda_\beta, \lambda_{\bar{\nu}_e})$ values at specific $(T, \rho Y_e)$ conditions. We evaluate the antineutrino spectral function as 
\begin{equation}
 n_i(E_{\bar{\nu}_e}) = \mathcal{N}_i E_{\bar{\nu}_e}^{2} (E_{\bar{\nu}_e} - q_i)^{2}
 \frac{e^{(E_{\bar{\nu}_e} - \mu_e - q_i)/T}}{1 + e^{(E_{\bar{\nu}_e} - \mu_e - q_i)/T} }
\Theta(q_i - E_{\bar{\nu}_e}), 
\end{equation}
with effective $Q$-value, $q_i$, determined by reproducing the average antineutrino energy $\langle E_{\bar{\nu}_e} \rangle = \lambda_{\bar{\nu}_e} / \lambda_\beta$, and $\mathcal{N}_i$ the overall normalization constant. The differential antineutrino emissivity, henceforth antineutrino emissivity, is subsequently given by
\begin{equation}\label{eq:emissivity}
    \frac{d \epsilon_{\bar{\nu}_e}}{ d\Omega dE_{\bar{\nu}_e}} = \sum\limits_i \frac{E_{\bar{\nu}_e}}{4\pi} Y_i n_i({E_{\bar{\nu}_e}}),
\end{equation}
where the summation is performed over all nuclei whose abundance $Y_i$ is given by the nuclear statistical equilibrium (NSE) composition. The nuclear composition is determined by the SFHo equation of state~\cite{Steiner_2013}. Employing our rates, the emissivities are calculated using the open-source neutrino library \texttt{NuLib}~\cite{O_Connor_2015}. The compilation of weak-decay rate tables for CCSNe simulations was pioneered in Refs.~\cite{Langanke2003,Hix2003}, and subsequently implemented in \texttt{NuLib} in Ref.~\cite{Sullivan_2016}, relying on analytic approximation where microscopic rates were unavailable. In this work, while we focus on $\beta$~decay rates, we employ the EC rates from a recent work (see Ref.~\cite{Ravlic2025}), obtained within the same theoretical framework. To study the impact of new $\beta$~decay rates on antineutrino emissivity, we examine the results obtained with three models. In the first case, we neglect the contribution of $\beta$~decays and only consider the antineutrino emission due to thermal rates. These include positron capture on free neutrons $e^++n \to p +\bar{\nu}_e$, electron-positron annihilation and nucleon-nucleon bremsstrahlung, with formalism described in \texttt{NuLib} documentation~\cite{O_Connor_2015}. For models that consider $\beta$~decays, we utilize our calculations (labeled FT-QRPA in the following) or SM calculations from Refs.~\cite{LANGANKE20011,Zha2019}, mostly in the $pf$-shell region. The CCSNe simulation is performed using the open-source code \texttt{GR1D}~\cite{O_Connor_2015} solving for early and post-bounce stages in spherical symmetry with general relativistic hydrodynamics. All simulations performed in this work consider either a 15$M_\odot$ or 20$M_\odot$ solar-metallicity progenitor (s15WW95, s20WW95)~\cite{1995ApJS..101..181W}.

\subsection{Competition Between Electron Capture and $\beta$ Decay}\label{sec:competition}
The EC is known as one of the most prominent weak-decay processes during the CCSNe collapse. However, we would like to assess thermodynamical conditions under which $\beta$-decay can compete with EC. To this aim, Fig.~\ref{fig:S1} shows the regions of the nuclide chart where $\beta$~decay and electron capture (EC) rates become comparable at representative thermodynamic conditions along the core-collapse supernova (CCSN) trajectory. While the $\beta$~decay rates are calculated in this work, the EC rates are taken from Ref.~\cite{Ravlic2025}, obtained within the same theoretical framework. For each nucleus, the two processes are classified as ``competitive’’ when the rates differ by less than one order of magnitude, \textit{i.e.}, $\left|\log_{10}\!\left(\lambda_{\mathrm{EC}} / \lambda_{\beta}\right)\right| \leq 1$.
Nuclei with $\lambda_{\mathrm{EC}}$ exceeding $\lambda_{\beta}$ by more than one order of magnitude are identified as EC dominated, while nuclei with $\lambda_{\beta}$ larger than $\lambda_{\mathrm{EC}}$ by more than one order of magnitude are labeled as $\beta$~decay dominated.

The full set of nuclei shown spans $20 \leq Z \leq 40$ and $16 \leq N \leq 82$, consistent with the domain of the FT-QRPA rate tables, including 902 individual nuclei. At relatively low temperatures and densities [Fig.~\ref{fig:S1}(a)], $\beta$~decay dominates much of the domain on the neutron-rich side, with slight competition near the valley of stability. At such low densities, $\beta$~decays are unhindered by phase-space blocking and dominate in neutron-rich nuclei due to the enhanced $Q_\beta$ window. On the other hand, for proton-rich nuclei, where $Q_\beta$ is low or even negative, EC dominates. In the in-between region, closer to the valley of stability, both rates can compete, though only over a very narrow band. We note that although most of these nuclei would be $\beta$-stable at zero temperature, finite-temperature effects can unblock the $\beta$~decay rates. As temperature and density increase during collapse [Figs.~\ref{fig:S1}(b)--(d)], the competitive band shifts to more neutron-rich nuclei and broadens toward the neutron-rich region, reflecting the fact that EC rates start increasing rapidly at higher temperatures and densities, while the $\beta$~decay phase-space gets Pauli blocked as the density is increased. Finally, in Fig.~\ref{fig:S1}(d), as the collapse progresses to hotter, more dense conditions, EC dominates fully.

\begin{figure*}
    \centering        
    \includegraphics[width=0.85\linewidth]{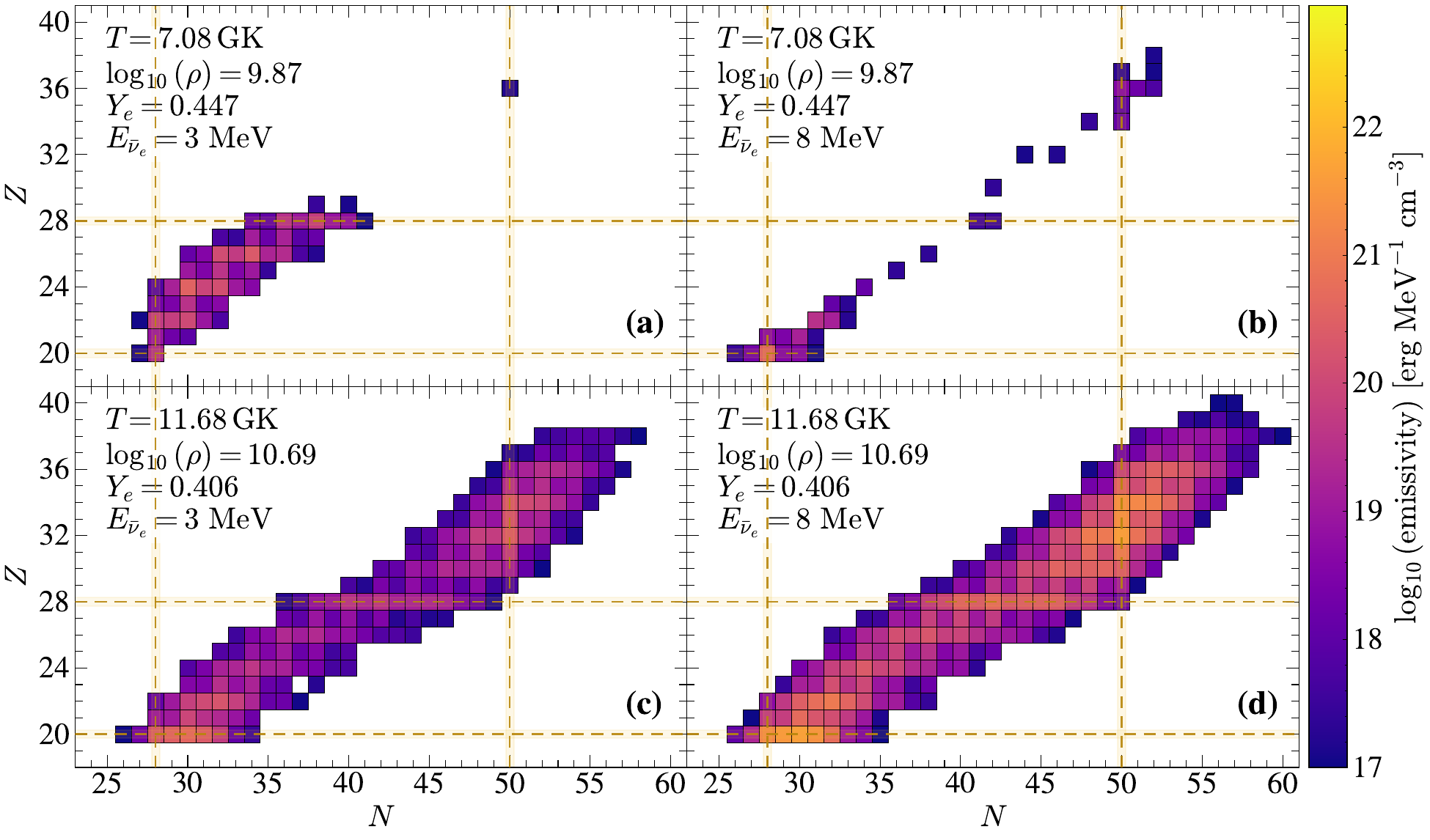}
    \caption{Partial contribution of individual nuclei to the total antineutrino emissivity at $E_{\bar{\nu}_e} = 3$~MeV (a,c) and $E_{\bar{\nu}_e} = 8$~MeV (b,d) at conditions corresponding to the onset of the collapse (a,b) and closer to the time of the bounce (c,d). All rates are obtained with the FT-QRPA model. Magic numbers are
indicated by dashed vertical and horizontal lines.}
    \label{fig:2}
    \hfill
\end{figure*}

We also note that the competitive regions identified in Fig.~\ref{fig:S1} partially coincide with the nuclei that contribute most strongly to the antineutrino emissivity (shown in Fig.~\ref{fig:2} and discussed in Sec.~\ref{sec:results}). Although the thermodynamic conditions are not identical, they lie along the same CCSN trajectory and probe similar density–temperature regimes. In particular, comparison with the emissivity patterns at comparable thermodynamic points demonstrates that the nuclei classified here as competitive, and often $\beta$~decay dominated, have a significant impact on the antineutrino emissivity at $E_{\bar{\nu}_e} = 3$~MeV and 8 MeV. This is especially the case for nuclei around the $N = 50$ region.

\subsection{Comparison of FT-QRPA and Shell-Model $\beta$~Decay Rates}\label{sec:comparison}

Figure~\ref{fig:S2} compares the $\beta$~decay rates obtained in this work with available shell-model calculations from Ref.~\cite{Zha2019} by showing the logarithmic ratio $\log_{10}\!\left(\lambda_{\mathrm{FT\text{-}QRPA}} / \lambda_{\mathrm{SM}}\right)$ along the CCSN trajectory.
Only nuclei for which both sets of rates exist are included, restricting the comparison primarily to the $pf$-shell region with $20 \le Z \le30$ and $22 \le N \le38$.

 Across all representative thermodynamic conditions shown, FT-QRPA rates are typically higher than shell-model values, even up to 3 orders of magnitude. This is consistent with the observations in Ref.~\cite{PhysRevC.104.054318} for $\beta$~decay rates obtained within a similar theoretical framework, as well as for the EC in Ref.~\cite{Ravlic2025}. Considering the fact that for $pf$-shell nuclei, the strength is mostly dominated by the allowed transitions, first-forbidden being negligible, as a possible reason for the observed differences, we note that considerable $\beta$-strength in the FT-QRPA originates from de-excitations, \textit{i.e.} transitions from the highly-excited states in the parent nucleus.

\begin{figure}
    \centering
    \includegraphics[width=0.9\linewidth]{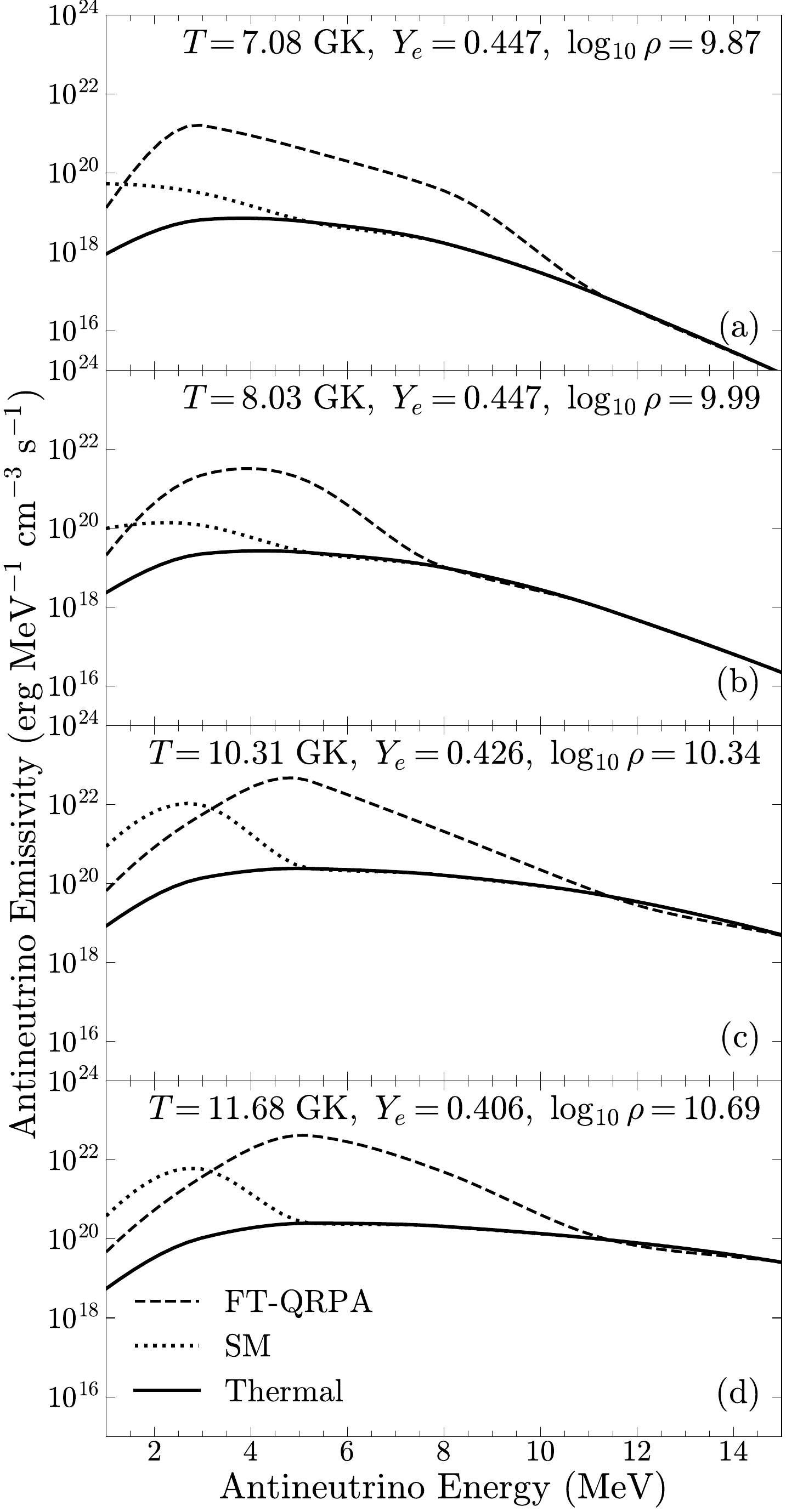}
    \caption{ The antineutrino emissivity as a function of antineutrino energy at 4 sets of thermodynamic conditions during the CCSNe evolution of a $20M_\odot$ progenitor (a)--(d). The emissivity due to thermal antineutrinos (solid line), as a baseline in all models, is compared to emissivities where the $\beta$~decay rates are explicitly included either from the FT-QRPA calculations in this work (dashed line) or the shell-model data (dotted line)~\cite{LANGANKE20011,Zha2019}.}
    \label{fig:1}
\end{figure}

\section{Results}\label{sec:results}
 Results for the antineutrino emissivity [cf. Eq. (\ref{eq:emissivity})] are shown in Fig.~\ref{fig:1} (panels (a)--(d)) across four representative points on the CCSNe thermodynamic trajectory of a $20 M_\odot$ progenitor. Emissivities obtained using three different models are compared. 
 Our baseline model, labeled Thermal, includes only contribution to antineutrino emissivity due to thermal processes, excluding nuclear $\beta$ decays. For the models containing nuclear $\beta$~decay, one model uses the FT-QRPA rates for 902 nuclei calculated herein, and the other uses SM rates from Refs.~\cite{LANGANKE20011,Zha2019}.
 Starting from the onset of the collapse in Fig.~\ref{fig:1}(a), we observe a huge enhancement of antineutrino emissivities due to inclusion of $\beta$~decays, relative to the baseline model. The FT-QRPA calculation yields an increase of more than two orders of magnitude, peaking around $E_{\bar{\nu}_e}\approx3$ MeV and extending up to $\approx 11$ MeV. For the SM rates, there is a lower enhancement, peaked at $E_{\bar{\nu}_e} \approx 1$~MeV, and converging rapidly to the thermal baseline for $E_{\bar{\nu}_e} \approx 5$ MeV. 
 The $\beta$~decay is constrained by the $Q_{\beta}$ energy window [cf. Eq.~(\ref{eq:beta_rate})]. This window measures approximately 11 MeV for neutron-rich nuclei but is significantly lower for $pf$-shell nuclei.
 We note that the SM rates are available for nuclei up to $A < 65$, which can only span a limited range in the $Q_\beta$ values. On average, as was remarked in Ref.~\cite{Ravlic2025}, the FT-QRPA rates are higher than those predicted by the SM as shown in Fig.~\ref{fig:S2}.
 This difference is attributable to multiple factors, such as the inclusion of first-forbidden transitions and the consistent treatment of de-excitations within the FT-QRPA. With increasing temperature, $\beta$~decay rates tend to increase, while with increasing density, the rates are significantly hindered~\cite{PhysRevC.104.054318}. We notice that with increasing temperature and density in Fig.~\ref{fig:1}(b)--(d) both FT-QRPA and SM emissivities stay approximately the same, with the peaks shifting to larger $E_{\bar{\nu}_e}$, primarily due to $\beta$~decay rates being blocked by increasing density, while the contribution of  thermal antineutrino baseline increases. Therefore, there is a window, up to around $\log_{10}\rho Y_e \approx 10$, where $\beta$~decay can produce a significant number of antineutrinos.

 \begin{figure*}
     \centering
     \includegraphics[width=0.7\linewidth]{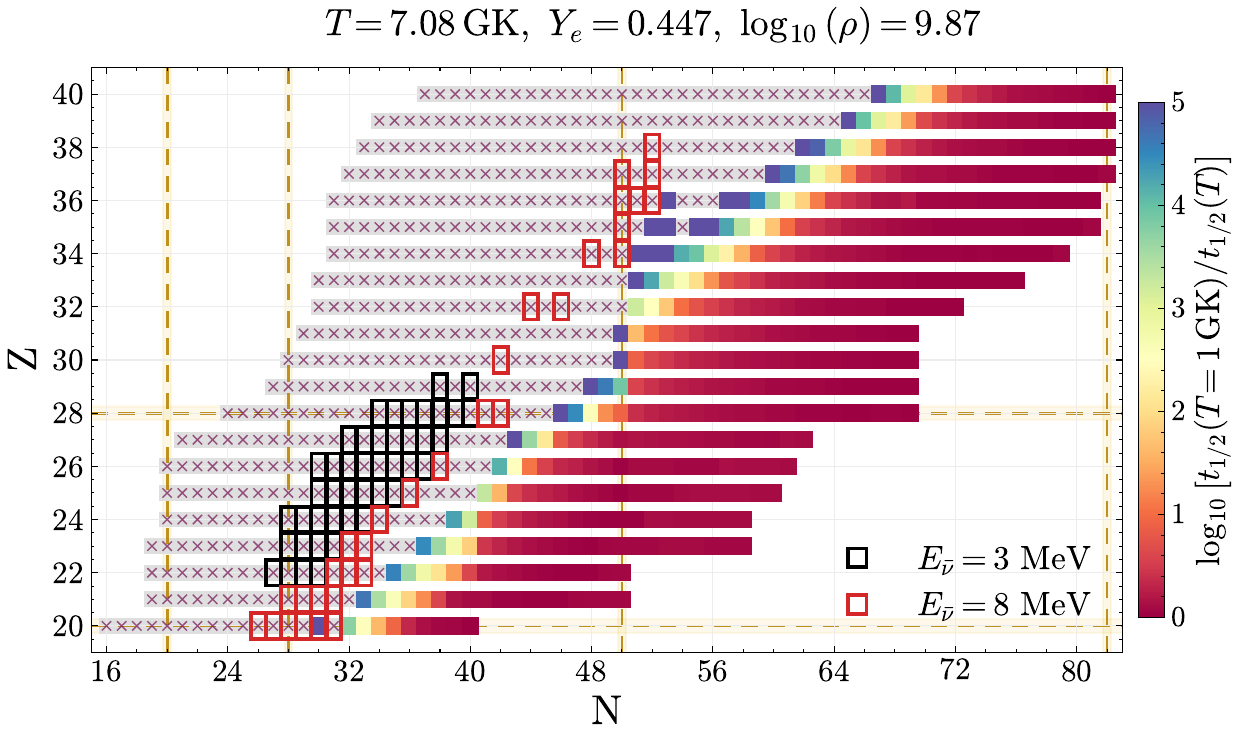}
     \caption{The ratio between $\beta$ decay half-lives $t_{1/2}$ at low temperature reference (taken as 1 GK) and $T = 7.08$ GK, at CCSNe conditions corresponding to $Y_e = 0.447$ and $\log_{10}\rho = 9.87$. Nuclei with infinite reference half-lives, corresponding to stability against $\beta$ decay in the low-temperature calculation, are marked with \textit{x} symbols. Important nuclei contributing to antineutrino emissivity under these conditions, from Fig.~\ref{fig:2}, are shown at $E_\nu = 3$ MeV (black) and $E_\nu = 8$ MeV (red). Magic numbers are indicated by dashed vertical and horizontal lines.}
     \label{fig:baseline_rates}
 \end{figure*}

 To study in greater detail which nuclei contribute the most to enhanced emissivity due to inclusion of $\beta$~decays, in Fig.~\ref{fig:2}, we show a \textit{partial} emissivity for individual nuclei, by restricting Eq.~(\ref{eq:emissivity}) to a specific nucleus $i$. Results are shown as obtained with the FT-QRPA $\beta$~decay rates. Based on the conclusions from Fig.~\ref{fig:1} we focus on two antineutrino energies, $E_{\bar{\nu}_e} = 3$ MeV and $E_{\bar{\nu}_e} = 8$ MeV. 
 Figure~\ref{fig:2}(a) identifies the nuclei primarily responsible for the emissivity peak observed in Fig.~\ref{fig:1}(a).
 Predictably, most of these nuclei are $pf$-shell. However, since the enhancement of the rates extends all the way to $E_{\bar{\nu}_e} \approx 8$ MeV, Fig.~\ref{fig:2}(b) displays the nuclei that contribute most significantly to the emissivity at the same conditions, specifically constrained to $E_{\bar{\nu}_e} = 8$ MeV. As expected, we observe a more extended distribution of nuclei that contribute, even up to the $N = 50$ region, with larger $Q_\beta$ values, demonstrating that even at lower temperature and density conditions the FT-QRPA tabulation is needed. Moving on, at higher temperatures and densities, closer to the conditions corresponding to those at the time of neutrino trapping, as seen in Fig.~\ref{fig:1}(d), an even more extended distribution of nuclei contributes to the total emissivity. However, in Fig.~\ref{fig:2}(d), we can see two regions of the nuclide chart that display a more enhanced contribution to the emissivity located around $(Z, N) = (20, 30)$, as well as the neutron-rich $(Z, N) = (32, 50)$. An increase in $E_{\bar{\nu}_e}$ from 3 to 8 MeV enhances contribution of nuclei around $N = 50$ significantly as shown in Fig.~\ref{fig:2}. 
 Therefore, predictive modeling of antineutrino spectra requires rate libraries that extend well into the medium-heavy mass region, such as those generated by our FT-QRPA calculations.

 To assess the role of temperature-induced unblocking on $\beta$ decay rates explicitly, Fig.~\ref{fig:baseline_rates} compares the $\beta$ decay half-lives $t_{1/2}$ at thermodynamic conditions corresponding to panels (a,b) of Fig.~\ref{fig:2}, with the low-temperature case (taken to be 1~GK in the current table). The baseline for the ratio is taken at the same $\log_{10} \rho Y_e = 9.52$. At such a high density, $\beta$ decay is mostly blocked due to high degeneracy of the lepton phase space. However, finite-temperature effects can thermally unblock those rates under these conditions. In Fig.~\ref{fig:baseline_rates} we observe that a large number of nuclei are denoted in gray, which means that at 1~GK they are stable and the ratio becomes infinite. Interestingly, almost all nuclei that contribute to antineutrino emissivity at $E_\nu = 3$ MeV and 8 MeV are found within this region. Beyond this region, the impact of temperature becomes progressively weaker. While nuclei closer to the valley of stability show greater acceleration with temperature, those further away and closer to the neutron drip line remain almost constant. This is consistent with predictions in Ref.~\cite{Ravlic2021}. For the higher-density conditions corresponding to panels in Fig.~\ref{fig:2}(c,d), $\log_{10} \rho Y_e = 10.30$, nearly all nuclei are blocked in the low-temperature limit, including the most important nuclei highlighted in Fig.~\ref{fig:2}(c,d). Again, temperature effects are essential for unblocking the $\beta$ decay rates.

\begin{figure}[htbp]
    \centering        
    \includegraphics[width=\linewidth]{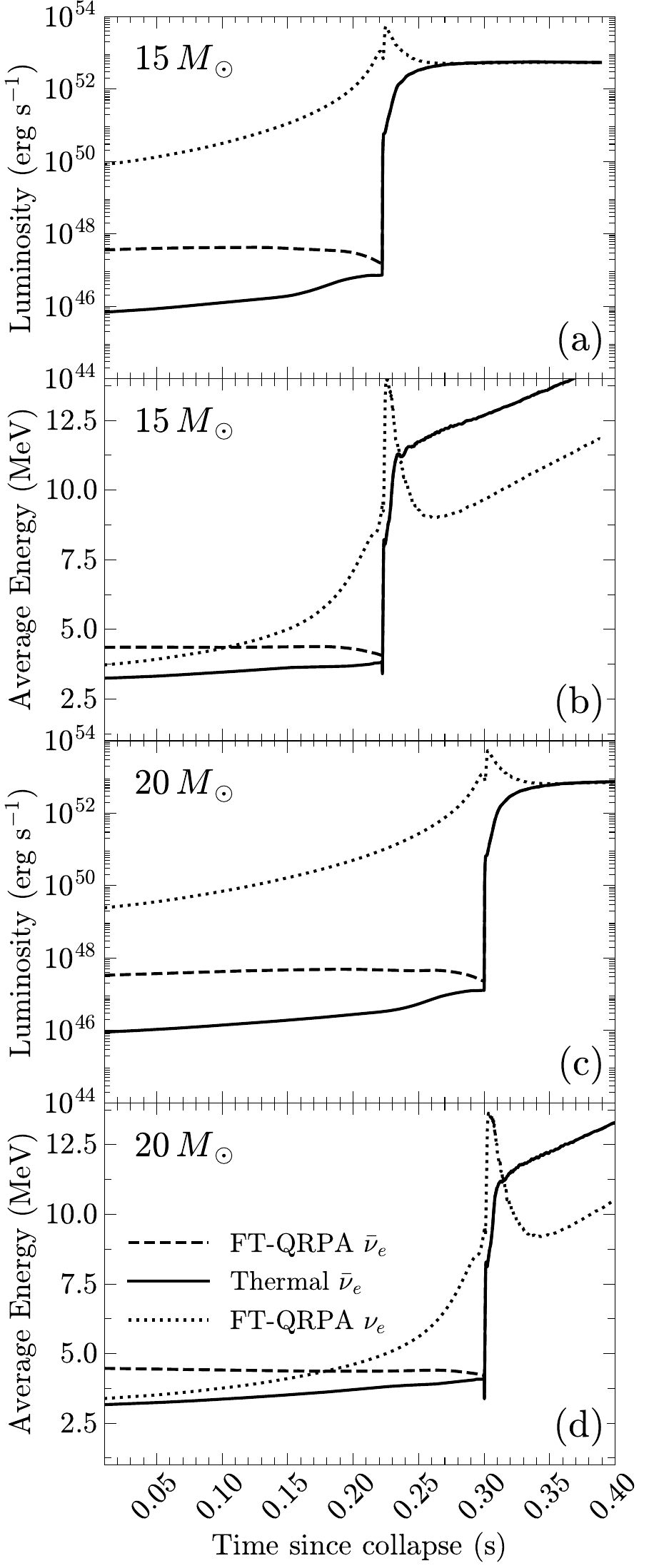}

    \caption{The time-evolution of the (anti)neutrino luminosity at 500 km from the core (a,c) and the average (anti)neutrino energy (b,d) for 15$M_\odot$ (a, b) and 20$M_\odot$ (c, d) progenitors. Results are displayed for the thermal baseline (solid line) together with the predictions of the FT-QRPA  neutrinos (dotted line) and antineutrinos (dashed line). }\label{fig:3}
    \hfill
\end{figure}

 To study the impact of these rates, we implement them in CCSNe simulations using the EC rates from Ref.~\cite{Ravlic2025}. Panels (a) and (c) of  Fig.~\ref{fig:3} show the time evolution of the neutrino and antineutrino luminosity at 500 km from the core of $15M_\odot$ and $20M_\odot$ progenitors. The thermal baseline antineutrino luminosity is shown by solid lines, while the dashed and dotted curves show the (anti)neutrino luminosity obtained using the FT-QRPA $\beta$~decay rates. For the neutrino luminosity (dotted curves), the dominant component originates from the EC and shows exponential increase towards the bounce, which occurs at $t \approx 225$~ms ($15M_\odot$) and $t \approx 300$~ms ($20M_\odot$) after the onset of the collapse. The thermal antineutrino luminosity exhibits a gradual increase up to the bounce. Once the $\beta$~decays are accounted for, the antineutrino luminosity shows a considerable enhancement of around a factor of 50 compared to the thermal baseline. $\beta$~decays bring the antineutrino luminosity closer to within 2 orders of magnitude of the neutrino one, for $20M_\odot$ progenitor. However, since the neutrino luminosity increases very steeply, as EC on heavier nuclei becomes temperature-unblocked, we observe no impact on CCSNe dynamics itself. Increased luminosity of antineutrinos enhances the antineutrino signal originating from $\beta$~decays just prior to the bounce. To better showcase this, panels (b) and (d) in Fig.~\ref{fig:3} display the time-evolution of the average (anti)neutrino energy. We observe that the antineutrino energy remains stable up to the bounce, as expected, considering the limited $Q_\beta$ window and emissivity trends in Fig.~\ref{fig:1}. Similar trends are observed irrespective of the progenitor mass. Compared to the thermal baseline, the average antineutrino energy by including the FT-QRPA $\beta$~decays increases, by around 1.1(1.3) MeV for $15M_\odot$($20M_\odot$) progenitor, driven by the changes we observed in emissivity in Fig.~\ref{fig:1}. While the average antineutrino energy remains almost constant up to the bounce, the average electron neutrino energy continues to increase considerably, reflecting the rapid increase in EC rates with temperature and density, which drives further deleptonization.
 We note that this represents the first calculation of average neutrino energy using the FT-QRPA EC rates from Ref.~\cite{Ravlic2025}, providing a direct comparison between different (anti)neutrino channels with a consistent rate framework.

Certainly, results presented in Fig.~\ref{fig:3}, invite the question whether an enhanced antineutrino luminosity, as well as the average energy, could modify the antineutrino signal on Earth-based detectors. We use the SNEWPY package \citep{Baxter:2021,Baxter:2022} to assess the detectability of the increased electron antineutrino signal in the pre-bounce phase for each of the four models simulated in this paper located at a fiducial galactic supernova distance of 10\,kpc.  Within SNEWPY, the GR1D neutrino luminosity, average energy, and root mean squared energy for each species are used to generate a time series of pinched spectra, these are then processed via SNOwGLoBES \citep{Scholberg:2012} to predict the rate of inverse beta-decay (IBD; i.e. $\bar{\nu}_e+p \to n + e^+$) events using two different detector configurations, \texttt{wc100kt30prct} (scaled to 188kt to match the upcoming Hyper-K detector) and \texttt{scint20kt} (a JUNO-like detector configuration).  Within SNOwGLoBES, the efficiency for detecting low energy positrons (emitted in the IBD channel) is $\sim 0\%$ for positron energies $\lesssim 5\,$MeV in the \texttt{wc100kt30prct} configuration, mainly due to the low number of photons produced at these positron energies.  The efficiency for the \texttt{scint20kt} configuration is taken to be 100\%, owing to liquid scintillator's lower detection threshold.  Detailed predictions, especially given the low average energy of our neutrino signals, would require better and specific detector characterisations.  For each model we present in Table~\ref{tab:prebounce_ibd} the expected number of detected IBD events within the detectors as predicted by SNEWPY during the simulated time ($\sim 300\,$ms before core bounce).

The expected counts in the final 300\,ms are vanishingly small at 10\,kpc and in the best case scenario only approach $\mathcal{O}(1)$ at a distance of $\sim\,0.3$ kpc.  Nevertheless, the predicted rate does sensitively depend on the $\beta$ decay framework used.  A full prediction would require a longer pre-collapse evolution.

\begin{table}[h]                                                                                                                      
    \centering                                                         \caption{Predicted pre-bounce IBD event counts at $d = 10$~kpc, assuming no neutrino oscillations. Hyper-K-like rates are scaled from a 100~kt water-Cherenkov detector (30\% PMT coverage; \texttt{wc100kt30prct}) to the 188~kt Hyper-K    
  fiducial mass. JUNO-lilke rates use the \texttt{scint20kt} SNOwGLoBES configuration. Employed models are $15M_\odot$ progenitor with (\texttt{s15\_beta}) and without (\texttt{s15\_nobeta}) $\beta$-decay rates included in GR1D simulation, and likewise for $20M_\odot$ progenitor.}\label{tab:prebounce_ibd}
  \vspace{0.3cm}
    \begin{tabular}{llcc}                                              
        \hline\hline                                                               
          Model & Progenitor & Hyper-K-like  & JUNO-like  \\               
        \hline                                                                     
        \texttt{s15\_nobeta} & $15\,M_\odot$ & $2.28\times10^{-4}$ & $1.01\times10^{-4}$ \\                                                                                                                         
        \texttt{s15\_beta}   & $15\,M_\odot$ & $1.50\times10^{-3}$ & $1.61\times10^{-3}$ \\                                                                                                                         
        \texttt{s20\_nobeta} & $20\,M_\odot$ & $4.30\times10^{-4}$ & $2.01\times10^{-4}$ \\                                                                                                                         
        \texttt{s20\_beta}   & $20\,M_\odot$ & $2.40\times10^{-3}$ & $2.54\times10^{-3}$ \\                                                      
          \hline\hline                                                             
      \end{tabular}                                                               
\end{table}

\section{Conclusion}
 In this work, we identify the impact of $\beta$~decays in stellar environments on CCSNe antineutrino observables from large-scale, microscopic calculations.
The new FT-QRPA calculations span a much larger set of nuclei and produce a significantly enhanced antineutrino spectrum compared to SM calculations, which are mostly limited to $pf$-shell nuclei. In particular, the inclusion of more neutron-rich nuclei enhances the antineutrino spectrum by approximately two orders of magnitude relative to the thermal baseline. This enhanced spectrum extends almost up to $E_{\bar{\nu}_e} = 11$~MeV, primarily because the late-collapse emissivity is dominated by neutron-rich nuclei with $N \approx 50$. When incorporated into CCSNe simulations, the enhanced $\beta$~decay emissivity translates into a substantially larger ($\sim 50$ times) pre-bounce antineutrino luminosity and around 1~MeV increase in the average antineutrino energy, while leaving the collapse dynamics unchanged. 
Nonetheless, the current absence of impact during the collapse dynamics does not imply that $\beta$~decays are unimportant for stellar evolution, and consequently CCSNe dynamics. Due to $\beta$~decay's competition with EC at relatively lower densities and temperatures, these new rates have the potential to alter the star's evolution in the presupernova stage, which would in turn modify the initial conditions that shape CCSNe dynamics. A full investigation of these evolutionary effects requires dedicated stellar-evolution calculations and will be pursued in a future work. Within the present CCSNe simulation framework, starting just $\sim200$ milliseconds before the bounce, we demonstrate no observable antineutrino signal detectability on Earth-based detectors in the pre-bounce phase. However, future studies which start earlier in stellar evolution could potentially show enhanced antineutrino detectability due to new FT-QRPA $\beta$ decay rates.

\section{Data availability}
The $\beta$~decay rate table used in this work is available in Ref.~\cite{ascsn_rpx_github}.

\section{Acknowledgments}
Discussions with Remco Zegers, Simon Giraud, Jim Kneller and Kelly Patton are gratefully acknowledged. This work was supported  by the U.S. Department of Energy under Award Nos. DOE-DE-NA0004245 (NNSA, the Stewardship Science Academic Alliances program), DE-SC0023688, DE-SC0023128 (Office of Science, Office of Nuclear Physics), and DE-SC0023175 (Office of Science, NUCLEI SciDAC-5 collaboration) and by the Swedish Research Council (Project No. 2020-00452). This work benefited from support in part by the National Science Foundation under Grant No. OISE-1927130 (IReNA). Computational resources were provided in part by the Institute for Cyber-Enabled Research at Michigan State University. This work used TAMU ACES at TAMU through allocation PHY250120 from the Advanced Cyberinfrastructure Coordination Ecosystem: Services \& Support (ACCESS) program, which is supported by U.S. National Science Foundation grants No. 2138259, 2138286, 2138307, 2137603, and 2138296.

%

\end{document}